\documentclass[twocolumn,preprintnumbers,superscriptaddress,amssymb,prl,aps,10pt]{revtex4-1}

\usepackage{graphicx}
\usepackage{dcolumn}
\usepackage{bm}
\usepackage{color}
\usepackage{amsmath,amsfonts,amssymb}  
\usepackage{graphicx}
\usepackage[english]{babel}
\usepackage{booktabs,longtable}
\usepackage{natbib}

\newcommand{\abs}[1]{\left|{#1}\right|}

\begin{document}

\author{M.~Cubukcu}
\affiliation{Unit\'e Mixte de Physique CNRS/Thales and Universit\'e Paris-Sud 11, 91767 Palaiseau, France}
\author{J.~Sampaio}
\altaffiliation[Present address: ]{Laboratoire de Physique des Solides, Univ. Paris-Sud, CNRS UMR 8502, 91405 Orsay Cedex, France}
\affiliation{Unit\'e Mixte de Physique CNRS/Thales and Universit\'e Paris-Sud 11, 91767 Palaiseau, France}
\author{K.~Bouzehouane}
\affiliation{Unit\'e Mixte de Physique CNRS/Thales and Universit\'e Paris-Sud 11, 91767 Palaiseau, France}
\author{D.~Apalkov}
\affiliation{Samsung Electronics, Semiconductor R\&D Center (Grandis), San Jose, CA 95134, USA}
\author{A.~V.~Khvalkovskiy}
\altaffiliation[Present address: ]{Moscow Institute of Physics and Technology, State University, Moscow, Russia}
\affiliation{Samsung Electronics, Semiconductor R\&D Center (Grandis), San Jose, CA 95134, USA}
\author{V.~Cros}
\affiliation{Unit\'e Mixte de Physique CNRS/Thales and Universit\'e Paris-Sud 11, 91767 Palaiseau, France}
\author{N.~Reyren}
\email{nicolas.reyren@thalesgroup.com}
\affiliation{Unit\'e Mixte de Physique CNRS/Thales and Universit\'e Paris-Sud 11, 91767 Palaiseau, France}

\title{A Dzyaloshinskii-Moriya Anisotropy in Nanomagnets with in-plane Magnetization}

\date{\today}

\begin{abstract}
We report on a new source of in-plane anisotropy in nanomagnets due to the presence of Dzyaloshinskii-Moriya interaction (DMI). This anisotropy depends on the shape of the magnet, and is orthogonal to the demagnetization shape anisotropy. This effect originates from the DMI energy reduction due to an out-of-plane tilt of the spins at edges oriented perpendicular to the magnetization. Our investigation combining experimental, numerical and analytical results demonstrate that this energy reduction can compensate the demagnetization shape anisotropy energy in magnets of elongated shape, provided that their volumes is small enough and thus that their magnetization is quasi-uniform.
\end{abstract}

\maketitle

The Dzyaloshinskii-Moriya interaction (DMI), which is the antisymmetric analog of the Heisenberg exchange energy, reduces the energy of a magnetic system when neighboring spins $\vec{S}_i$ and $\vec{S}_j$ are not parallel, and can be expressed in an Hamiltonian as $\vec{d}_{ij}\cdot\left(\vec{S}_i \times \vec{S}_j \right)$, where $\vec{d}_{ij}$ is the Dzyaloshinskii-Moriya vector. The DMI is known to be a direct manifestation of spin-orbit coupling (SOC) in systems with broken inversion symmetry \cite{Dzyaloshinskii1958,Moriya1960}. During the last couple of years, DMI has triggered an increasing interest and a lot of research in magnetism community because it could lead to several technological breakthroughs in the area of magnetic recording. In particular, the DMI can impose the structure of magnetic domain walls (DW) in perpendicular magnetic films to change from conventional Bloch into chiral N\'eel configurations \cite{Thiaville2012,Chen2013,Tetienne2015}, controlling the direction of propagation and increasing dramatically the velocity of DW in racetrack structures \cite{Emori2013,Ryu2013,Parkin2015}. Interestingly, this chiral interaction can also lead to new magnetic textures, such as spin spirals or magnetic skyrmions of a few atomic lattice periodicity/size \cite{Muhlbauer2009,Yu2010} that hold promises as a new approach for magnetic memories \cite{Kiselev2011,Fert2013,Sampaio2013}.

Large interfacial DMI of magnitude comparable with the Heisenberg exchange can be brought about in thin metallic multilayers made of $3d$ ferromagnetic ({\it e.g.} Fe, Co, Ni) and $5d$ heavy metals ({\it e.g.} Ta, W, Ir, Pt), in which the inversion symmetry is broken at interfaces and the large SOC is provided by the heavy metal in contact with the ferromagnetic layer \cite{Meckler2009,Muhlbauer2009,Emori2013,Fert2013,Ryu2013}. The impact of the DMI in magnetic materials with out-of-plane magnetization were experimentally observed by several means, such as the measurement of the asymmetric growth and propagation of magnetic domains or DW with external magnetic fields \cite{Emori2013,Je2013,Hrabec2014,Hiramatsu2014,Pizzini2014}, the tilting of the magnetic DW in the presence of an external transverse magnetic field \cite{Boulle2013}, the local measurements of the magnetization or its stray field using scanning probes \cite{Ferriani2008,Tetienne2015}, the study of the spin wave propagation in magnetic multilayers with Brillouin light spectroscopy \cite{Belmeguenai2015}, or the observation of magnetic skyrmions by spin-polarized STEM or scanning transmission x-ray microcopy with polarized light \cite{Chen2015,Jiang2015,Moreau-Luchaire2015}. Most of these techniques and observations rely on the particular structure of the DW induced by the DMI, namely a N\'eel configuration in which the chirality (sense of rotation of magnetization along an axis lying in the DW plane) is fixed by the sign of the DMI. A DMI-induced tilt of the magnetization close to the edge of out-of-plane magnetic structures has also been predicted \cite{Rohart2013,Sampaio2013}, but there are no experimental evidences of this phenomenon yet.

For symmetry reasons, most of the studies of the DMI have been performed with perpendicularly magnetized ultra-thin films. In this letter, we investigate the effect of the DMI in magnetic films with in-plane magnetization. We demonstrate the existence of an original in-plane anisotropy term due to the tilt of the magnetization that is induced by the DMI at edges of magnetic nanostructures \cite{Rohart2013}. We first present experimental observations of the unexpected magnetization direction in sub-micrometer-scale ellipses made of a thin layer of in-plane magnetized CoFeB sandwiched between a Pt and an MgO layer: For the smallest observed structures, the magnetization lies along the minor axis of the ellipses. This behavior is well reproduced in micromagnetic simulations in which the interfacial DMI is considered. This observation also allows the micromagnetic DMI amplitude $D$ to be indirectly determined, providing that other micromagnetic parameters are known. We then explain the origin of the effect and give an approximated analytical formulation for the critical $D$ value, $D_\perp$, that brings the magnetization from lying along the major axis to the minor axis in rectangular prisms. Finally, we present a set of phase diagrams of the magnetic configurations as a function of these micromagnetic parameters.

\begin{figure}
\includegraphics[width=8.5cm]{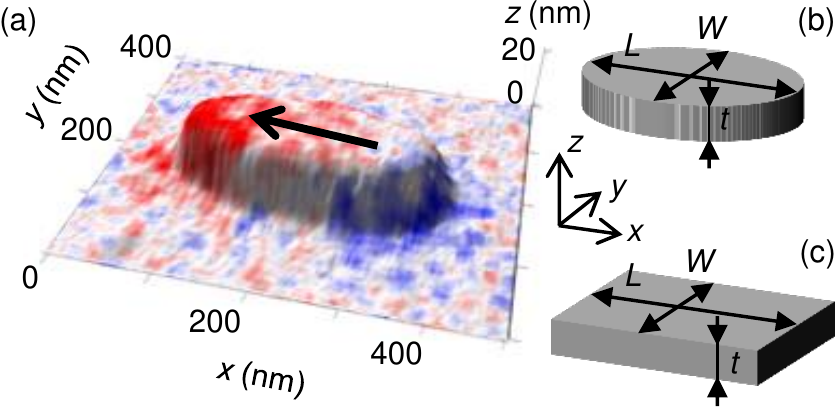}
\caption{(Color online) AFM/MFM measurement and geometry of the problem. (a) Actual measurement showing in pseudo-3D the topographical signal (AFM) colored with the phase of the MFM signal. The MFM phase is coded blue or red for attractive or repulsive forces, allowing the mean magnetization to be determined, as indicated by the arrow. (b,c) Geometries used to perform micromagnetic numerical simulations and the analytical calculations (c).}
\label{Fig1}
\end{figure}

The studied samples are composed of MgO$|$Co$_{20}$Fe$_{60}$B$_{20}$\,(1.5\,nm)$|$Pt\,(0.8\,nm) stacks that are grown by sputtering deposition. We fabricated elongated nanostructures (nominally ellipses) with variable areas and aspect ratios, using e-beam lithography and Ar ion beam etching. The major axis of the studied ellipses nominally ranges from 1\,$\mu$m to 100\,nm, and their aspect ratios from 1 to 10 [Fig.\,\ref{Fig1}(b,c)]. We have used atomic and magnetic force microscopy (AFM/MFM) to image both the topographies (real sizes and shapes) and the magnetic configurations (direction of the magnetization) of these nanostructures. Particular attention should be taken about the choice of a tip with ultra-low moment to avoid perturbing the sample magnetization by the tip’s stray field. A typical elliptical structure is displayed in Fig.\,\ref{Fig1}(a) in which the topographical signal (AFM) is rendered in pseudo-3D and the MFM phase signal is color-coded.

\begin{figure}
\includegraphics[width=8.5cm]{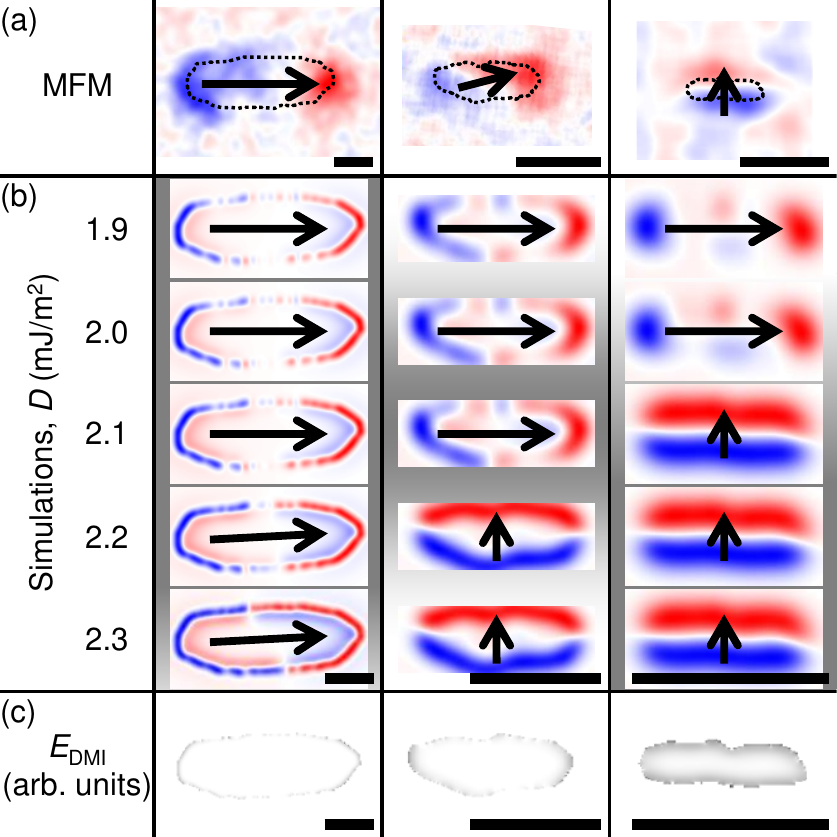}
\caption{(Color online) DMI-induced anisotropy orthogonal to the shape anisotropy as observed by MFM and simulations. All scale bars are 200\,nm long. The arrows indicate the mean magnetization. (a) Experimental MFM phase images are displayed. The actual topographical shape is indicated by a dotted line. The MFM phase is coded in the same way than in Fig.\,\ref{Fig1}. (b) Corresponding simulations for different $D$ values (all the other micromagnetic parameters fixed), using an ideal dipolar MFM tip 30\,nm above the magnetic layer. The observations of the magnetization direction in ellipses of different shapes (but same materials) allow constraining the Dzyaloshinskii-Moriya interaction strength $D$: The shadings indicate the $D$ values compatible with observations. In the intermediate case, the magnetization does not align along one of the axes, reflecting the minute energy difference between the states with different magnetization directions. (c) Map of the DMI and exchange energy densities (darker means lower energy density) for the case $D=2.1$\,mJ/m$^2$. The energy gain concentrates close to the edges.}
\label{Fig2}
\end{figure}

The experimental measurement of a series of ellipses of different sizes is presented in Fig.\ref{Fig2}(a). It illustrates how the magnetization turns from the major axis for large ellipses with $L\approx 0.7\,\mu$m (left MFM image) to the minor axis for a small ellipse with $L\approx 160$\,nm (right MFM image). In the intermediary case we found one ellipse ($L\approx 250$\,nm) with the magnetization pointing neither along the minor nor the long axis. As we will see below, this state is probably a limit case where the DMI-induced anisotropy equals the shape anisotropy. We emphasize that such effects were not observed on a symmetric reference sample made of MgO$|$Co$_{20}$Fe$_{60}$B$_{20}$\,(2.9\,nm)$|$MgO.

\begin{figure}
\includegraphics[width=8.5cm]{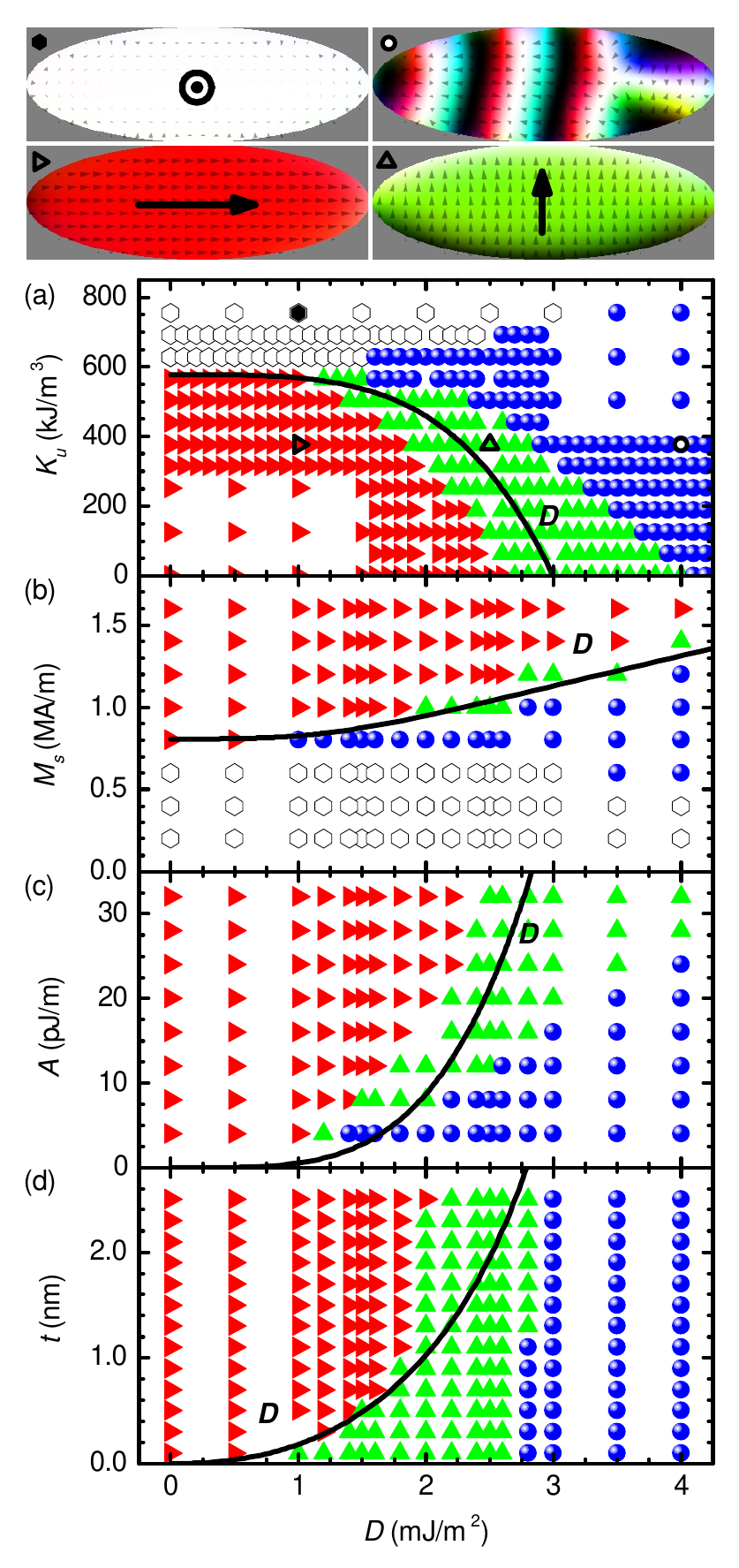}
\caption{(Color online) Numerical and analytical phase diagrams. Simulated state (ellipses) of for different $K_u$ (a), $M_s$ (b), $A$ (c) and $t$ (d) as a function of $D$, and the analytical critical $D_\perp$ for rectangles. Right pointing triangles (red) correspond to states with the magnetization along $x$, up triangles (green) along $y$, spheres (blue) are used if the state is very inhomogeneous with several domains, and hexagons (open) if the state is with perpendicular magnetization. The physical parameters of the simulations are $M_s=1$\,MA/m, $K_u=377$\,kJ/m$^3$, $A=15$\,pJ/m, $L=210$, $W=70$ and $t=1.5$\,nm if not indicated otherwise. Four typical micromagnetic states of panel (a) which correspond to the black symbols are drawn on top of the figure.}
\label{Fig3}
\end{figure}

We reproduce this behavior using micromagnetic simulations, for which we used the GPU-based MuMax3 code (version 3.6.1 and 3.8) which natively includes the interfacial DMI \cite{Vansteenkiste2014}. In our simulations, as the simulated systems are thinner than the exchange length, we consider that the magnetization is uniform over the thickness. The system is hence modeled with a single layer, and consequently discretization cells have their size along $z$ determined by the thickness $t$, while sizes of 0.4 to 0.8\,nm were considered along $x$ and $y$ directions. In order to define the geometrical shape of the simulated structure, we have used the exact sample dimensions determined by the AFM characterization (taking in account the effect of the tip width). The magnetic parameters used for the simulations are the magnetization at saturation $M_s=1.06\pm0.08$\,MA/m and the crystalline perpendicular anisotropy $K_u=0.26\pm0.06$\,MJ/m$^3$  both determined by SQUID measurements \cite{Sawicki2011}, and an exchange stiffness $A$ of $10$\,pJ/m typically found in the literature for this material \cite{Yamanouchi2011}. The value of the micromagnetic DMI parameter $D$ is still not available for a large range of interfaces, but a rather large value is expected at the interface between Pt and ferromagnetic materials \cite{Chen2013,Emori2013,Ryu2013,Hrabec2014,Pizzini2014,Belmeguenai2015}.  Simulations were hence performed for different values of $D$ and, as displayed in Fig.\,\ref{Fig2}(b), there is a change of magnetization direction as the DMI increases. Comparing simulations with MFM measurements allows the estimation of $D\approx 2.1_{-0.3}^{+0.4}$\,mJ/m$^2 $ at the CoFeB$|$Pt interface. This value can be compared, for example, to the Brillouin light scattering experiments where $D=1.2$\,mJ/m$^2$ for the Co(1.2\,nm)$|$Pt interface \cite{Belmeguenai2015} or 0.8 for CoFeB(0.8\,nm)$|$Pt \cite{Di2015}, or to domain wall experiments in Co(0.6\,nm)$|$Pt where $D=2.2$\,mJ/m$^2$ \cite{Pizzini2014}. It is important to note that the value of $D$ is very sensitive to the actual magnetic parameters of the material (as readily visible on the phase diagrams of Fig.\,\ref{Fig3}, {\it e.g.} $D\approx 2.4$\,mJ/m$^2$ if $A=15$\,pJ/m is used). However, provided a precise determination of the micromagnetic parameters ($M_s$, $K_u$ and $A$), the observation of nanoscale ellipses could hence lead to a novel and more direct way to estimate $D$. One should finally note that the energy minimization of the simulations must be studied with caution because the energy landscape is extremely flat close to the transition. This also explains why, in the real system, minute inhomogeneity might stabilize the magnetization along any direction (see for example the intermediate case of Fig.\,\ref{Fig2}).

In order to get a better insight into the mechanisms at play, an analytical model can be formulated. In magnetic thin films with moderate DMI, the DMI induces a tilt of the magnetization at the border. In perpendicularly magnetized films, this tilt occurs along the whole border of the structure \cite{Rohart2013}, while for in-plane magnetized films with interfacial induced DMI, the tilt only occurs at the borders with an orthogonal component of the magnetization. In particular, it is important to notice that there is no tilt if the magnetization is parallel to the edge, while it is maximal for a magnetization direction perpendicular to the border. For the latter case, the tilt of magnetization reduces the system energy by \cite{Heide2008,Rohart2013}:
\begin{equation*}
E_{\rm tilt} = - \frac{D^2}{4\sqrt{A K_{yz}}}  t L\quad ,
\end{equation*}
where $L$ is the border length, $t$ is the thickness of the magnetic film. The parameter $K_{yz}$ is the effective planar anisotropy constant and is expressed as $K_{yz}=\frac{1}{2} \mu_0 M_s^2 (\mathcal{N}_z-\mathcal{N}_y)-K_u$ for a simple shape with a remaining crystalline perpendicular anisotropy $K_u$, and $\mathcal{N}_i$ are the demagnetizing factors. As the borders where tilting occurs depend upon the direction of the in-plane magnetization, this phenomenon induces a planar anisotropy that competes with the dipolar-induced shape anisotropy. 

To describe this effect, let us consider a simple shape: a rectangular prism with dimensions $L$, $W$ and $t$ [along the $x$, $y$, and $z$ axes respectively as shown in Fig.\,\ref{Fig1}(c)]. We consider a system small enough to be in a single magnetic domain state. The energy difference $E_x-E_y$ between the states with the magnetization along $x$ and $y$ is then given by:
\begin{eqnarray*}
E_x-E_y&=&\Delta E_{\rm demag}+\Delta E_{\rm tilt} \\
&=&K_{xy} LWt-\frac{D^2 t}{4\sqrt{A}}\left(  \frac{2W}{\sqrt{K_{xz}}} - \frac{2L}{\sqrt{K_{yz}}}\right)\\
&\approx&\left(K_{xy}+\frac{D^2}{2\sqrt{A K_{xz}}} \frac{L-W}{LW}\right)LWt\quad ,
\end{eqnarray*}
where $K_{xy}=\frac{1}{2} \mu_0 M_s^2 (\mathcal{N}_x-\mathcal{N}_y)$  is the in-plane shape anisotropy and $K_{xz}=\frac{1}{2} \mu_0 M_s^2 (\mathcal{N}_z-\mathcal{N}_x)-K_u$. This DMI-induced tilt results in an in-plane anisotropy of direction orthogonal to the shape anisotropy, {\it i.e.} favoring the magnetization parallel to the smaller side of the structure. In the case of a rectangular parallelepiped, the ``DMI-anisotropy'' can be approximated as:
\begin{equation*}
K_{\rm DMI}\approx\frac{D^2}{2\sqrt{AK_{xz}}} \frac{L-W}{LW}  \quad .
\end{equation*}
The threshold value $D_\perp$ at which the magnetization switches from one axis to the other is then:
\begin{equation}
D_\perp^2 \approx (\mathcal{N}_x-\mathcal{N}_y)  \frac{WL}{W-L} \mu_0 M_s^2 \sqrt{A K_{xz}}\quad ,
\label{EqDperp}
\end{equation}
which is composed of a part that depends on shape and volume and another part that depends almost only on the material magnetic parameters. 

To illustrate this effect more comprehensively, in Fig.\,\ref{Fig3}, using micromagnetic simulations, we map out phase diagrams for elliptical shapes [see Fig.\,\ref{Fig1}(b)] of the magnetization direction as a function of the magnetic parameters $K_u$, $M_s$, $A$ and the geometrical parameter $t$ using typical parameters found for thin magnetic films. In each phase diagram, three regimes that are controlled by the DMI are observed: (1) at low DMI, the magnetization lies along the major axis as expected in classical systems, (2) at larger values of DMI, the DMI-anisotropy overcomes the shape anisotropy and aligns the magnetization along the minor axis, and (3) for very large DMI, the magnetization is not anymore single domain, but multi-domains, helices or skyrmions appear, because the presence of domain walls minimizes the energy, as each of these DW, being of N\'eel type, reduces the DMI energy. Illustrations of typical configurations are given in the top part of Fig.\,\ref{Fig3}. The tilt at the edges perpendicular to the magnetization are also clearly visible (color-coded black and white for down or up magnetization).
 
The analytical model (Eq.\,\ref{EqDperp}) can be used to plot $D_\perp$ in a simplified case where the demagnetization factors of the rectangular prism are considered for a uniform magnetization \cite{Aharoni1998}. As shown in Fig.\,\ref{Fig3}, the analytical curves are reproducing the trends of the micromagnetic simulated system. The difference between $D_\perp$ in the micromagnetic simulations (the border between the red right triangles with the green up triangles) and the analytic curve cannot be simply explained by the $\mathcal{N}_i$ of the different shapes ({\it i.e.} demagnetizing factors for rectangles and ellipses). Indeed, simulations were also realized for rectangular nanomagnets showing a behavior much closer to the ellipses than to the analytical model. The comparison of the different components of the energy indicates that the main limitation of the analytical model (which uses a strictly uniform magnetization) is that the demagnetization energy is very much reduced by the tilting occurring at the edges, hence reducing the needed DMI-anisotropy necessary to align the magnetization along the minor axis/short side of the structures. Nevertheless, the simple analytical model allows a very quick assessing of the potential role of the DMI-anisotropy in a system.

\begin{figure}
\includegraphics[width=8.5cm]{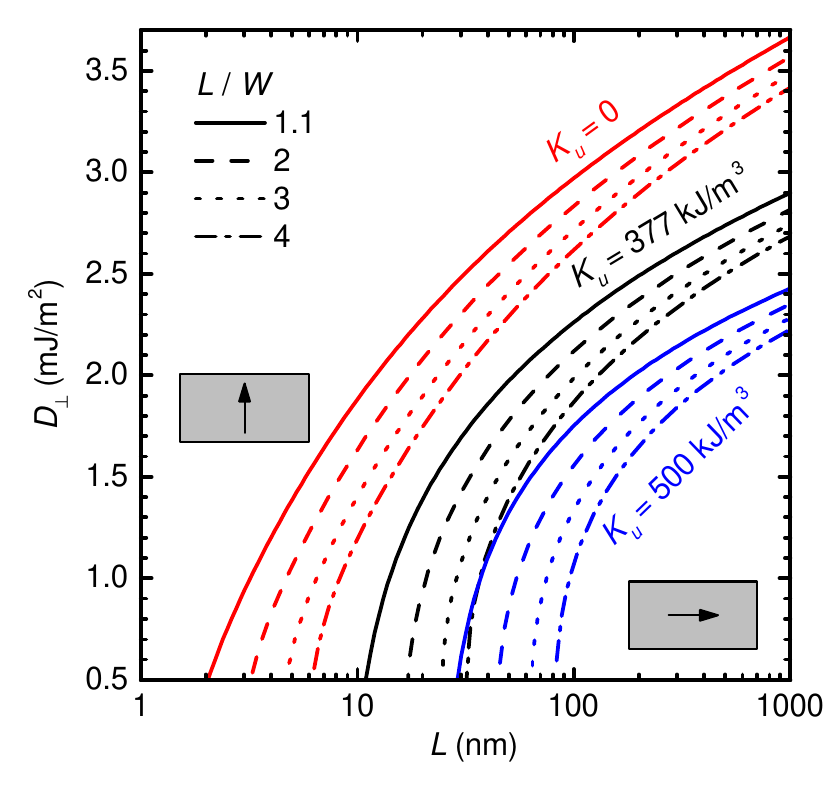}
\caption{(Color online) Analytical $D_\perp$, critical $D$ for transverse magnetization as a function of the long size of a rectangle for different aspect ratio and $K_u$. The uniaxial perpendicular anisotropy plays a crucial role and diminishes greatly $D_\perp$.}
\label{Fig4}
\end{figure}

In Fig.\,\ref{Fig4}, the dependence of the analytical critical $D_\perp$ for transverse magnetization is plotted as a function of the length ($L$) for different aspect ratios ($L/W$) and $K_u$. A larger value of the out-of-plane anisotropy $K_u$ favors the DMI-anisotropy, because the tilting at edges increases (in absolute value) the associated negative energy term. As expected, $D_\perp$ decreases with increasing aspect ratio, and smaller structures require weaker DMI to induce the transverse stabilization of the magnetization. In particular, it is interesting to note that nanostructures of the size of typical MRAM cells will undoubtedly be sensitive to this effect if Pt is in contact to the magnetic layer, as a DMI as high as 0.5\,mJ/m$^2$ are certainly expected.

Finally, we note that the energy difference between states with magnetization along $x$ or $y$ reaches values greater than 1\,eV, when far enough from the $D_\perp$ line on the phase diagram ({\it e.g.} $\abs{D-D_\perp}\geq 0.2$\,mJ/m$^2$ in the case of the ``standard'' nanomagnet of Fig.\,\ref{Fig3}). Such energy difference indicates a good stability against temperature of the states induced by the DMI-anisotropy, as $\abs{E_x-E_y}/(k_B T)\approx 40$ for 1\,eV at 300\,K.

In conclusion, we observe and explain an original ``DMI-anisotropy'' for in-plane magnetized nanomagnets. We show that, for small volumes and reasonable aspect ratios, a strong DMI-associated energy is taking over the shape anisotropy demagnetization energy, and that the magnetization is aligning along the minor axis/short side of elongated structures. We can note that the energy difference between these states is of the order of the electron-volt, meaning that they are thermally very stable. We also present MFM observation constituting the first experimental evidence of the tilting of the magnetization at the edges of magnetic structures due to DMI. This phenomenon used in giant magnetoresistance or tunnel magnetoresistance structures could possibly be used to electrically detect small magnetic field with a device of extremely small size that maybe useful, for example, in detectors or magnetic read heads.

\begin{acknowledgments}
This work was supported by the Samsung Global MRAM Innovation Program. The authors would like to thank S.~Xavier from Thales TRT for the e-beam lithography.
\end{acknowledgments}


%

\end{document}